\documentclass[preprint,bm,showpacs]{revtex4}

\usepackage{graphicx}
\usepackage{amssymb}

\def\ba{\begin{eqnarray}}
\def\ea{\end{eqnarray}}
\def\be{\begin{equation}}
\def\ee{\end{equation}}
\def\ben{\begin{equation} \nonumber}
\def\een{\end{equation}}
\def\ban{\begin{eqnarray*}}
\def\ean{\end{eqnarray*}}

\def\Na{N_{\alpha}}

\def\D{\overline{\mbox{D}}}
\def\nalpha{n_\alpha}
\def\nbeta{n_\beta}
\def\ngamma{n_\gamma}
\def\Nalpha{N_\alpha}
\def\Nbeta{N_\beta}
\def\Ngamma{N_\gamma}
\def\mualpha{\mu_\alpha}
\def\falpha{f_\alpha}

\def\rhoa{\rho_\alpha}

\def\baray{\begin{eqnarray}}
\def\earay{\end{eqnarray}}

\begin{document}
\title{Scaling of Multi-Tension Cosmic Superstring Networks}
\author{S.-H. Henry Tye$^{1}$} \email{tye@lepp.cornell.edu} 
\author{Ira Wasserman$^{1,2}$}\email{ira@astro.cornell.edu}\author{Mark Wyman$^{1,2}$}\email{wyman@astro.cornell.edu}
\affiliation{$^1$ Laboratory for Elementary Particle Physics,
Cornell University,
Ithaca, NY 14853, USA \\
$^2$ Center for Radiophysics and Space Research, Cornell
University, Ithaca, NY 14853, USA }

\begin{abstract} 

Brane inflation in superstring theory ends when branes collide, 
initiating the hot big bang. Cosmic superstrings are produced 
during the brane collision. The cosmic superstrings produced in
a D3-brane-antibrane inflationary scenario have a spectrum: 
$(p,q)$ bound states of $p$ fundamental (F) strings and $q$ D-strings, 
where $p$ and $q$ are coprime. By extending the velocity-dependent
one-scale network evolution equations for abelian Higgs cosmic strings 
to allow a spectrum of string tensions, we construct a coupled (infinite) 
set of equations for strings that interact through binding and self-interactions. 
We apply this model to a network of $(p,q)$ superstrings. Our numerical solutions 
show that $(p,q)$ networks rapidly approach a stable scaling solution. 
We also extract the relative densities of each string type from our solutions.  
Typically, only a small number of the lowest tension states
are populated substantially once scaling is reached.
The model we study also has an interesting new feature: the energy
released in $(p,q)$ string binding is by itself adequate to allow the network to
reach scaling. This result suggests that the scaling solution is robust.
To demonstrate that this result is not trivial, we show that choosing 
a different form for string interactions can lead to network frustration.
\end{abstract}

\maketitle

\section{Introduction}      

The cosmic microwave background (CMB) data from WMAP \cite{cobe,wmap} 
strongly supports the inflationary universe \cite{guth} as the 
origin of the hot big bang. 
Recently, the brane world scenario suggested by superstring theory
was proposed, where the standard model of the strong and electroweak 
interactions are open string (brane) modes while the graviton and the 
radions (which govern the size/shape of the bulk) are closed string (bulk) modes. 
Consider a generic brane world scenario that closely describes our 
universe today (i.e., a KKLT-like vacuum \cite{Kachru:2003aw}). 
If we take an extra brane-anti-brane pair in the early universe
their brane tensions provide the cosmological constant that drives 
inflation. There is an attractive force between a brane and an 
anti-brane so they tend to move towards each 
other while inflation is taking place. 
Thus, brane inflation is a natural feature of the brane 
world \cite{dvali-tye,collection,Kachru:2003sx}; in brane inflation, 
the inflaton is an open string mode identified with the interbrane separation.
Inflation ends when the D3-$\D$3-brane pair collides and annihilates, releasing energy that 
starts the hot big bang. 
Note that the inflaton field no longer exists after the annihilation of the 
D3-$\D$3-brane pair. 
Towards the end of inflation, the  D3-$\D$3-brane collision produces D1-branes -- or 
cosmic superstrings -- but neither monopoles nor domain walls \cite{Jones:2002cv}.

We can estimate the cosmic string tension $\mu$ using the density 
perturbation magnitude in the CMBR data from COBE \cite{cobe}. 
In the simplest realistic scenario, namely, the KKLMMT 
D3-$\D$3-brane inflationary scenario \cite{Kachru:2003sx}, one finds that \cite{Firouzjahi:2005dh} 
\baray
	 5 \times 10^{-7} \ge G \mu \ge 4 \times 10^{-10}
\earay
where $G$ is the Newton's constant. The upper bound comes from WMAP 
data and other data \cite{Pogosian:2003mz, us05}, while the lower values require some fine-tuning
in the model. These predictions are not sensitive to different warping schemes because the
normalization to COBE is always performed after the warp effect is taken into account. In this scenario,
the density perturbation responsible for structure formation is dominated by the inflaton, 
with cosmic strings playing a secondary role.

The evolution of networks of cosmic strings is a well studied problem \cite{Vilenkin,stringnet}.
After the initial production of cosmic strings, the strings interact among themselves. 
When two cosmic strings intersect, they reconnect or intercommute. 
When a cosmic string intersects itself, a closed string loop is broken off. 
Such a loop will oscillate quasi-periodically and gradually lose energy via 
gravitational radiation. Its eventual decay transfers the cosmic string 
energy to gravitational waves. A higher (lower) string density leads to a 
higher (lower) interaction rate so, not surprisingly, cosmic string 
networks evolve towards scaling solutions. A consequence of scaling is that the 
physics of simple, abelian Higgs networks is essentially dictated by a single parameter, 
the dimensionless string tension $G \mu$. This scaling feature can be seen by considering 
the evolution of the number density in a one-scale model, where the scaling solution 
emerges as an attractive fixed point.

There are many different hybrid inflationary models in which one can construct 
a variety of cosmic string-producing scenarios. What is different in brane inflation
is that the string network that is produced has a large spectrum of 
possible string tensions \cite{giaalex,Copeland:2003bj,Jones:2003da}.
For the D3-$\D$3-brane inflationary scenario, one expects a spectrum of
$(p,q)$ string bound states \cite{giaalex, Copeland:2003bj}, where the tension of a 
particular bound state $(p,q)$ is given by
\be
\label{Gmu1}
\hspace{2cm}G\mu_{(p,q)} = G \mu \sqrt{p^2g^2_s + q^2}.
\ee
where $g_s$ is the superstring coupling. For the string to be stable, $p$ and $q$ must 
be coprime, i.e., $p$ and $q$ have no common factors greater than $1$.
Written this way, $(1,0)$ corresponds to the fundamental F-string while $(0,1)$ corresponds
to the $D$1-brane, or D-string, so $\mu$ is the tension of the $(0,1)$ superstring. 
We can see the emergence of such a spectrum in several ways.
Consider the following simple picture:
the gauge group at the end of inflation just 
before brane-antibrane annihilation is 
$$
U(1)_D \times U(1)_{\bar D} = U(1)_{+} \times U(1)_{-},
$$
where the open string (complex) tachyon field stretching between the branes
couples only to $U(1)_{-}$. The usual operation of the Higgs mechanism generates
abelian vortices following the spontaneous symmetry breaking of $U(1)_{-}$.
These are the D-strings.
Since no free $U(1)$ gauge symmetry remains after the annihilation,
it is believed that the  $U(1)_{+}$ symmetry becomes confining, yielding confining fluxes
that may be identified as fundamental closed strings, or F-strings \cite{Bergman:2000xf}.
The production of D-strings may be estimated via the Kibble mechanism. 
Most of the decay products are expected to be very massive non-relativistic
closed strings, which are expected to decay to gravitons, standard model 
particles and other light modes. We expect some of the massive closed strings
to be extended. These are the F-strings. In a cosmological setting, 
their production is likely, again, to be dictated by the Kibble mechanism. 
The production of D- and F-strings are not independent, so we expect some
initial spectrum of $(p,q)$ strings to be produced. 

Since the interactions between $(p,q)$ strings is not simple \cite{Jackson:2004zg}, one
expects that $(p,q)$ network evolution might be quite involved.
It is not obvious,\emph{a priori}, that the network can even approach scaling. 
For example, it could oscillate (i.e., the density of any specific
$(p,q)$-type could oscillate indefinitely), approach scaling only asymptotically, 
or simply frustrate.
One way to address this problem would be to do a full numerical simulation of a 
$(p,q)$ string network.  However, this is a highly non-trivial problem.
String network evolution is a complex physical process; accurately modeling the build-up of 
small-scale string structure
is computationally demanding, even in the context of 
abelian Higgs models, which have only one type of vortex. 
A radically simpler alternative would be simply to generalize the 
one-scale string network model due to Kibble \cite{Kibble} to the case of $(p,q)$ string evolution. 
Recall that the scaling of the cosmic string network appears as a stable fixed point in this one-scale model. However, previous researchers have found it useful to include more of the network physics
than the original one-scale model allowed.
In particular, the velocity-dependent one-scale (VOS) model developed by Martins and Shellard 
for the abelian Higgs case \cite{Martins} provides a very convenient and reliable method for 
calculating the large-scale quantitative properties of 
string networks in many contexts, including a cosmological setting. 
This model performs exceptionally well when tested against
high resolution numerical simulations of string networks \cite{stringnet}. 
It allows one to see analytically how scaling emerges, 
and to calculate reliably a small number of macroscopic quantities 
useful for cosmological applications. We take this model as the starting
point for our own model building.
We recognize that there are a number 
of other analytic approaches to the  string evolution \cite{allen}. 
Some may characterize the details of small-scale stringy structures more accurately, but they also require more phenomenological input parameters which can only be obtained from simulations.
 Since there is, as yet, no simulation of a cosmic superstring network, and since 
 we are chiefly interested in the overall properties of such a string network,
 we choose the simplest possible ``analytic" model that highlights
 the most important physical effects.
 
In this paper we adapt the velocity-dependent one-scale model to describe
a multi-tension cosmic string network that includes both string self-intersection
and string-string binding interactions. For a multi-tension network,
the string density evolution equation generalizes to a set of (infinitely many) coupled equations. 
We then specify the string interaction terms in our particular multi-tension, 
$(p,q)$ string model and solve this set of equations numerically. 
Fortunately, we find that the $(p,q)$ string network, with stringy interactions
turned on, rapidly approaches 
scaling. This fast convergence and rapid decrease of string
densities with increasing tension allows us to truncate the set of equations at 
low, computationally tractable values of $(p,q)$. 
To show that this scaling result is not somehow built into the model 
we have constructed, we demonstrate that the same set of evolution 
equations with a different interaction term can lead to a frustrated network. 

Because of the various approximations we use, our results are limited to overall 
macroscopic network features, which are nonetheless those features that 
are needed for cosmological applications.
For instance, we find for the $(p,q)$ superstring network :
\begin{itemize}

\item The $(p,q)$ string network approaches a scaling network rapidly.
The final scaling solution is independent of the initial densities of the various types of strings.
 The fractional density in strings, for $F \neq 0$, is given by 
 \be
  \Omega_{cs}  = 8  \pi G \mu_{(0,1)} \Gamma  \quad \quad \Gamma \simeq \left \{ \begin{array}{cr} 20 / (0.55 P + F) &~~~ g_s  = 1.0 \\
  15/(0.53P + F) &~~~ g_s = 0.5
  \end{array} \right.
\ee
where $P$ measures the probability of self-interaction and $F$ measures the overall probability
of interaction among different types of strings. For the $(p,q)$ cosmic superstring network, we 
do not know the value of $F$, though we expect $P \lesssim F \lesssim 1$.
It is interesting to note that scaling is achieved even if we turn off the string self-interaction, i.e., 
when $P=0$.
For the abelian Higgs model we have $\Gamma_{U(1)} \simeq 20$ and $P\simeq 0.28$.
Using this value for $P$ and taking $F=P$, we find that, for $g_s=1$, $\Gamma \simeq 46$;
for $F=1$, $P=0.28$, we have $\Gamma \simeq 17$. Thus, the total density of the $(p,q)$
cosmic superstring network is comparable to standard cosmic strings.
Differentiating between the two kinds of networks based on their string densities 
will require more detailed modeling. 

\item The relative number density of each type of string is roughly given by
\be
N_{(p,q)} \sim \mu_{(p,q)}^{-n} \;\;\;\;~~ 6  < n \lesssim10.
\ee
The fall-off is power-like, not exponential. The rapid convergence of the coupled set of 
equations is brought about by this rapid fall-off. The power law is most accurate for high values
of $p$ and $q$; the spectrum tends to be somewhat flatter for the first few string types.
Indeed, we find that when scaling is reached, the relative numbers of $(0,1)$, $(1,0)$,
and $(1,\pm1)$ strings are comparable and far larger than the population of the
remaining $(p,q)$ states with $p,\, |q| > 1$. In the case of $F=P=0.28$, $g_s = 1.0$,
we find $N_{(p,q)} \propto \mu_{(p,q)}^{-7.5}$.

\item The adapted multi-tension velocity-dependent one-scale model (MTVOS) that we 
describe in \S \ref{VOSmulti} can be used for many different kinds of multi-tension networks 
by a simple change of the inter-string interactions term.

\end{itemize}

Clearly, this analysis can be improved in a variety of ways. We shall comment on some of them. However, we are confident that the rapid approach to scaling and the fast power drop-off in the 
densities are generic features of cosmic superstring networks. 

In the past, the evolution of cosmic strings in models more complicated than the abelian
Higgs model have been considered. For example, Pen and Spergel
studied the evolution of $S_N$ strings by simulating a network of $S_3$
and $S_8$ strings \cite{penspergel}. The $S_N$ symmetry enforces 
identical tension and number densities among the $N$ string types.
In terms of the model we construct, the set of $N$ coupled evolution
equations collapses to a single equation. McGraw also modeled 
non-abelian $S_3$ strings with two different tensions \cite{mcgraw}. 
The evolution of networks of $Z_N$ strings connected to monopoles, which 
have some qualitative similarities to the networks we consider,
has been studied in Ref. \cite{VV}.
We believe that a network of $(p,q)$
strings is the first network type that truly requires a set of coupled equations.
The formalism may be adapted for other non-trivial string network.

In Section 2, we adapt the velocity-dependent one-scale model to a 
model for the evolution of comic strings that have a spectrum of tensions. 
In Section 3, we specialize the model to the $(p,q)$
superstring network by defining our string interaction term.
In Section 4 we present our numerical results and in Section 5 we briefly
discuss some observational implications of these networks.

\section{The Multi-Tension Velocity-Dependent One Scale Model}
\label{VOSmulti}

Consider a set of different types of cosmic strings
$\{\alpha\}$ with tensions $\{\mu_\alpha\}$. Let
the number  of cosmic strings of type
$\alpha$ per unit area  be $\nalpha$. Suppose that all of the cosmic
strings may be characterized by a single length scale $L$ and 
a single average velocity $v$, and that cosmic strings of type $\alpha$ can evolve 
either by interaction mediated loop formation or by binding to cosmic
strings of other types $\beta\neq\alpha$. The following model 
is motivated by the model of Martins and Shellard \cite{Martins},
but has been altered somewhat to accommodate the new string physics
we introduce.

We assume that the length scale evolves via the equation
\be
\dot L=HL+c_1v~,
\label{Leqn}
\ee
where the loop parameter $c_1\leq 1$ is a dimensionless factor and $H$ is the Hubble parameter. 
We take the equation of motion for the velocity to have
the Martins-Shellard form
\be
\dot v=(1-v^2)\left(-2Hv+{c_2\over L}\right)~;
\label{veqn}
\ee
where the ``momentum parameter" $c_2$ is a second constant. 
This term is the acceleration due to the curvature of the strings.
In the absence of expansion, these two equations imply
$\gamma=(1-v^2)^{-1/2}=\left({L/ L_0}\right)^{c_2/c_1}$,
and with $c_1=c_2=0$, but with expansion retained, they imply
$\gamma va^2={\rm constant},$ 
${L/ a}={\rm constant}$, and so, $\gamma v L^2={\rm constant}.$
Thus, the ``self-acceleration'' due to string curvature 
and expansion have opposite effects: self-acceleration
increases string velocity, whereas expansion dilutes it.
This suggests that the two effects can cancel one another.
We can see how this comes about by rewriting Eq. (\ref{Leqn})
as
\be
{d(HL)\over dt}=H\left(HL+c_1v\right)+{\dot H\over H}HL
=H(HL)\left[{c_1v\over HL}-\left({1+3w\over 2}\right)\right]~
\ee
and combining it with the velocity equation;
we then find that there is a quasi-steady solution
\be
v=HL\left({1+3w\over 2c_1}\right)={c_2\over 2HL}~,
\ee
which implies
\be
HL=\sqrt{c_2c_1\over 1+3w},~~~~~~~~~
v={1\over 2}\sqrt{c_2(1+3w)\over c_1}~.
\ee
In this solution, both $HL$ and $v$ are constants that differ
in the radiation ($w=1/3$) and matter ($w=0$) eras. Clearly, there is no
quasi-steady solution for $w\leq -1/3$; thus, quasi-steady solutions
only exist in the radiation and matter eras. We require that in both
eras, $v\leq 1$, a condition that is more restrictive in the
radiation era, where it demands that $c_2/2c_1\leq 1$.
In practice, we choose $c_1$ and $c_2$ such that the scaling 
values of $HL$ and $v$ match the
values given in \cite{Martins}, where similar constants $\tilde{c}$ and $k$ are
chosen to line up with full network simulations. The translation
between our constants and theirs is simple: The translation
between our constants and theirs is simple: in the radiation era, 
our $c_2 = k$, while our $c_1 = (k+\tilde{c})/2)$; 
in the matter era, $c_2 = (3/4) k$ and $c_1 = (3/8)(k + \tilde{c})$.
 For example, in the radiation era they find $HL = 0.1375$ and $v=0.655$, 
which for us fixes $c_1= 0.21$ and $c_2 = 0.18$.

Next, add to these equations an equation of energy conservation, 
at first in the absence of interactions between strings of different types.
Let the cosmic string energy density be
\be
\rho={n\mu\over\sqrt{1-v^2}}
\label{rhodef}
\ee
where $\mu$ is the mass per unit length of a string, and $n$
is the mean string number density. In the absence of interactions,
assume that
\be
\dot\rho = -2H\rho(1+v^2)
\label{rhodot}
\ee
Differentiating the expression, Eq. (\ref{rhodef}) for the string
energy density and using Eq. (\ref{rhodot}) and Eq. (\ref{veqn}), the equation for $\dot v$,
implies
\be
\dot n=-\left(2H+{c_2v\over L}\right)n~.
\ee
where we see that the first term comes from cosmological expansion. The
second term can be interpreted in two different ways. For straight
strings, it would reflect a net expansion in the velocity field orthogonal
to the strings. For kinky strings, we should interpret $n$ as the
characteristic number of intersections per unit area on average for
a two dimensional surface intersecting the network. As the strings
straighten out (the source term in $\dot v$) the number of intersections
will fall, at a rate that is $\sim nv/L$ characteristically.
We may interpret this to mean that as string kinks straighten out,
the number of intersections of strings with
any two dimensional cut through three dimensional space will
decrease; it is also the straightening of kinks that raises $v$.
Note that when the string velocity approaches its asymptotic
value $v=c_2/2HL$, the energy equation becomes
\be
\dot\rho=-\left(2H+{c_2v\over L}\right)\rho
=-H\left[2+{c_2(1+3w)\over 2c_1}\right]\rho~,
\ee
i.e. it assumes exactly the same form as the equation
for the number density of strings. This is sensible because
the energy per unit length per string $\mu\gamma\to{\rm constant}$
asymptotically.

Next, let us consider what happens when we allow interactions between
strings. Recall that we assume a single characteristic length scale, $L$, for all types of strings;
however, unlike the single-$\mu$ case, this length need not be related directly
to string density. We further assume that the different types of cosmic strings interact by binding,
first at a point and then zipping up to form a new cosmic string with the 
same length as the original two, which enforces equal lengths for all different 
kinds of strings. The zipping up takes a time $L$, but 
let us suppose that $L$ is small enough compared to the Hubble length that
we can regard the zipping up as instantaneous; $HL \sim 0.1$ is
good enough for our purposes. Let the equation for 
$\nalpha$ be
\be
\dot\nalpha+2H\nalpha=-{c_2\nalpha v\over L}
-{P\nalpha^2vL}+FvL\left[{1\over 2}\sum_{\beta,\gamma}
P_{\alpha\beta\gamma}\nbeta\ngamma(1+\delta_{\beta\gamma})
-\sum_{\beta,\gamma}P_{\beta\gamma\alpha}\ngamma\nalpha
(1+\delta_{\gamma\alpha})\right]~,
\label{popeqn}
\ee
where the first term on the RHS arises from the breaking off of loops
from individual undulating strings of type $\alpha$, the second term
arises from breaking off of loops after the collision of two strings
of type $\alpha$, and the third term arises from the zipping up of
two strings of different types that collide and bind. 

As an aside, we note that the term proportional to $P$ could equally well be 
written as a term under the summation, proportional to $P_{0 \alpha \alpha}$. 
We have pulled this term out of the sum to make the contrast between self interaction
and interaction between strings of different types clear. But a few comments 
on this term are in order:
\begin{itemize}
\item The term proportional to $P$ or $P_{0 \alpha \alpha}$ is the usual term that drives 
networks without multiple string types to scaling.
\item By writing $P$ instead of $P_{0 \alpha \alpha}$, we are assuming that the self-interaction rate
 does not depend upon $\alpha$.
\item If we take $P_{\alpha\beta\gamma} = 0$ for $\beta \neq \gamma$, then a self-interaction
term of this form drives all string species which have a sufficient initial density
to the same final scaling number density.
\item After we have taken $P$ out of the sum, we either restrict the sum to $\beta \neq \gamma$
or assume $P_{0 \alpha \alpha} = 0$.
\end{itemize}

We assume that our constants $c_2$, $P$, and $F$ are identical
for cosmic strings of all types. We define $F$ as a measure of the overall
probability that two strings of different types interact at all.
We have assumed that the interaction of strings of two different
types can only result in zipping up, if anything -- there are
no reconnections and no breaking off of loops directly associated
with such interactions.

In the third term, $P_{\alpha\beta\gamma}$ is the probability of forming
a string of type $\alpha$ when strings of types $\beta$
and $\gamma$ collide, whenever the strings interact at all. The factor 
$${1\over 2}\left(1+\delta_{\beta\gamma}\right)$$
is introduced so that we do not double count the production 
of strings of type $\alpha$ when strings of {\it different}
types $\beta$ and $\gamma$ collide; i.e.
since we do not restrict the sum, symmetry on $\beta\leftrightarrow
\gamma$ implies we have two identical source terms from 
$\beta+\gamma\to\alpha$; the factor
$$1+\delta_{\gamma\alpha}$$
in the loss rate arises because in each $\alpha-\alpha$ collision
we lose {\it two} long strings. 

By defining $\Nalpha=a^2\nalpha$, we can rewrite  Eq. (\ref{popeqn})  as
\be
\dot\Nalpha=-{c_2\Nalpha v\over L}-{P\Nalpha^2vL\over a^2}
+{FvL\over a^2}\left[{1\over 2}\sum_{\beta,\gamma}
P_{\alpha\beta\gamma}\Nbeta\Ngamma
(1+\delta_{\beta\gamma})
-\sum_{\beta,\gamma}
P_{\beta\gamma\alpha}\Ngamma\Nalpha(1+\delta_{\gamma\alpha}
)\right]~,
\ee
and if we define conformal time by $d\eta=vdt/a$ and
introduce the comoving string length, $L=a\ell$, then
we find
\be
\Nalpha'=-{c_2\Nalpha\over\ell}-{P\Nalpha^2\ell}
+F\ell\left[{1\over 2}\sum_{\beta,\gamma}
P_{\alpha\beta\gamma}\Nbeta\Ngamma
(1+\delta_{\beta\gamma})
-\sum_{\beta,\gamma}
P_{\beta\gamma\alpha}\Ngamma\Nalpha(1+\delta_{\gamma\alpha})\right]~,
\label{etapopeqn}
\ee
with a prime denoting differentiation with respect to
conformal time. In terms of $\Nalpha$, we find that
$\rhoa=\mualpha\Nalpha/a^2\sqrt{1-v^2}$.

The remaining two equations are those for $\ell$ and $v$. 
Substituting $L=\ell a$ into Eq. (\ref{Leqn}) gives
\be
\dot\ell={c_1v\over a}~~\Rightarrow~~\ell=\ell(0)+c_1\eta~.
\ee
When we change the independent variable from $t\to\eta$,
Eq. (\ref{veqn}) becomes
\be
v^\prime={(1-v^2)\over v}\left(-2Hav+{c_2\over\ell}
\right)~.
\ee
To complete the set of equations, we need to find $a(\eta)$;
we use
\be
H={d(\ln a)\over dt}={d(\ln a)\over d\eta}{d\eta\over dt}
={v\over a}{d(\ln a)\over d\eta}
\ee
to get
\be
{da\over d\eta}={Ha^2\over v}~.
\ee
In the radiation dominated era, which is of greatest interest
to us practically, $Ha^2$ is approximately constant. Thus,
in the quasi-steady state, the scale factor grows linearly
with $\eta$.

Eq. (\ref{popeqn}) may yield a steady state solution of the
form $\Nalpha=\falpha/\ell^2=\falpha/[\ell(0)+c_1\eta]^2$
provided that
\be
-2c_1\falpha=-(c_2\falpha+P\falpha^2)+F
\left[{1\over2}\sum_{\beta,\gamma}P_{\alpha\beta\gamma}
f_\beta f_\gamma\left(1+\delta_{\beta\gamma}\right)
-\sum_{\beta,\gamma}P_{\beta\gamma\alpha}f_\gamma
f_\alpha\left(1+\delta_{\gamma\alpha}\right)\right]
\label{scaling}
\ee
has a nontrivial solution.

It is instructive to consider what we get when $F=0$. In 
that case, Eq. (\ref{scaling}) has two solutions, $\falpha
=0$, which is not relevant, and
\be
\falpha={2c_1-c_2\over P}~,
\label{onescale}
\ee
which is physically realizable only if $2c_1>c_2$. Let us
assume that this is so. At sufficiently late times, we
will therefore find that
\be
\rhoa={\mualpha(2c_1-c_2)\over Pc_1^2(a\eta)^2\sqrt{1-v^2}}~,
\label{rhoaone}
\ee
where we have assumed that $c_1\eta\gg\ell(0)$, and that
$v$ relaxes to its asymptotic value. In this limit, we
find that $a\approx Ha^2\eta/v$ in the radiation era,
and therefore $\eta\approx va/Ha^2=v/Ha$; use this in
Eq. (\ref{rhoaone}) to find
\be
\rhoa\approx{\mualpha H^2(2c_1-c_2)\over Pc_1^2v^2\sqrt{1-v^2}}
~~~\Rightarrow~~~\Omega_\alpha={8\pi G\rhoa\over 3H^2}
\approx {8\pi G\mualpha(2c_1-c_2)\over 3Pc_1^2 v^2\sqrt{1-v^2}}~.
\label{omalpha}
\ee
Because this version of the network equations assumes
common $L$ and $v$ for all string types, when $F=0$
we expect to find $\Omega_\alpha/\mu_\alpha$ independent
of $\alpha$, assuming nonzero initial populations.
(Remember that $\Omega_\alpha\propto\falpha=0$ is
also a solution.)
Notice that Eq. (\ref{omalpha}) implies $\Omega_\alpha
\propto(2c_1-c_2)/P$, and for small self-interaction
probability, this is the expected $P^{-1}$ scaling.
Moreover, nonzero $P$ is essential for time-independent
$\Omega_\alpha$ to arise
in networks with only interactions among strings of the
same type. Also, since this is the limit in which our model
reduces to the abelian Higgs case, we can use prior 
simulation results to fix the value of our parameter $P$.
Numerical studies of the radiation dominated era
have $\Gamma = \Omega_{cs} / (8\pi G\mu) \approx 20$, 
which for us implies $P =  0.28$, taking $c_1 = 0.21$,
 $c_2 = 0.18$, and $v = 0.655$. Because of this we take $P=0.28$
 as the fiducial value for $P$ in our numerical solutions.

It is also instructive to consider what happens if $P=F=0$.
In this case, Eq. (\ref{popeqn}) has an exact solution
\be
\Nalpha={\Nalpha(0)\ell(0)^{c_2/c_1}\over[\ell(0)+c_1\eta]^{c_2/c_1}
}\to\Nalpha(0)\left[{\ell(0)\over c_1\eta}\right]^{c_2/c_1}~
\label{noptwo}
\ee
(Eq. (\ref{noptwo}) is a special case of the general solution
$${1\over\Nalpha}=\left[{1\over\Nalpha(0)}-{P\ell^2(0)\over
2c_1-c_2}\right]\left[{\ell\over\ell(0)}\right]^{c_2/c_1}
+{P\ell^2\over 2c_1-c_2}~$$
that can be found for $F=0$, and the results of this and
the previous paragraph follow from appropriate limiting
cases of this solution.). From this result, we find that
\be
\Omega_\alpha\approx{8\pi G\mualpha\Nalpha(0)[\ell(0)]^{c_2/c_1}
\over 3(Ha)^{2-c_2/c_1}v^{c_2/c_1}\sqrt{1-v^2}}~.
\ee
In the radiation dominated era, when $Ha\propto a^{-1}$, we
find that $\Omega_\alpha\propto a^{2-c_2/c_1}$, which either
rises or falls depending on the sign of $2-c_2/c_1$.

Note that the network equation, Eq. (\ref{popeqn}), may also be used to describe
entanglement. In that case, instead of setting $\alpha=
(p_\alpha,q_\alpha)$, as we shall do to describe $(p,q)$ networks,
 we simply let $\alpha=p_\alpha$.
Moreover, we set $P_{\alpha\beta\gamma}=\delta_{\alpha-
(\beta+\gamma)}$; taking $c_2=P=0$, the network equations are
\be
\Nalpha'=F\ell\left[{1\over 2}\sum_{\beta,\gamma}
\delta_{\alpha-(\beta+\gamma)}\Nbeta\Ngamma(1+\delta_{\beta\gamma})
-\Nalpha\sum_\gamma\Ngamma(1+\delta_{\gamma\alpha})\right]~.
\label{entangle}
\ee
These equations cannot lead to a scaling solution because
there is a conservation law, basically conservation of energy,
that restricts the evolution of the system. Multiply
Eq. (\ref{entangle}) by $\alpha$ and sum over $\alpha$;
the result is
\ba
\left(\sum_\alpha\alpha\Nalpha\right)'&=&F\ell
\biggl[{1\over 2}\sum_{\alpha,\beta\neq\gamma}\alpha
\delta_{\alpha-(\beta+\gamma)}\Nbeta\Ngamma
-\sum_{\alpha\neq\gamma}\alpha\Nalpha\Ngamma\nonumber\\& &
+\sum_{\alpha,\beta}\alpha\delta_{\alpha-2\beta}
\Nbeta^2-2\sum_\alpha\alpha\Nalpha^2\biggr]
\nonumber\\
&=&F\ell\left[{1\over 2}\sum_{\beta\neq\gamma}(\beta+\gamma)\Nbeta
\Ngamma-\sum_{\alpha\neq\gamma}\alpha\Nalpha\Ngamma
+2\sum_\beta\beta\Nbeta^2-2\sum_\alpha\alpha\Nalpha^2\right]
\nonumber\\
&=&F\ell\left[\sum_{\beta\neq\gamma}\beta\Nbeta\Ngamma
-\sum_{\alpha\neq\gamma}\alpha\Nalpha\Ngamma
+2\sum_\beta\beta\Nbeta^2-2\sum_\alpha\alpha\Nalpha^2\right]
\nonumber\\
&=&0~,
\ea
where the next to last line was obtained by relabelling
$\gamma\to\beta$ in one of the two sums over $\beta\neq\gamma$.
Thus, here we have an example where scaling is {\it not} achieved.
The network evolves toward ever larger values of $\mu$, but its
overall comoving energy density does not decline.

Thus,  neither the existence nor the nature of a scaling solution 
for a particular multi-tension network is obvious.
If each type of string evolves independent of all other types,
scaling will be achieved eventually for all types present originally ,
with $\Omega_\alpha\propto\mualpha/P$. Turning on the interactions
between string types will populate the different tensions, and, once
produced, their self-interactions and energy-losing binding interactions
will propel them toward scaling
solutions. The final spectrum of string tensions may
be broad or narrow, depending on the efficiency with which the
reaction terms operate. The reaction terms themselves may promote
scaling even if there are no self-interactions, but the fact
that entanglement can be described by a reaction network with
particular choices of interaction probabilities shows that there
are certainly circumstances in which scaling cannot arise solely
from the reactions among strings of different types. Note, finally, that
we have assumed very little about the nature of the multi-tension
network that is described by these equations beyond the assumption
that string interactions lead either to loop formation or the 
formation of other kinds of strings through some sort of binding.
Thus these equations may easily be
adapted for any particular multi-tension string network model
simply by determining the form of $P_{\alpha \beta \gamma}$
for that model; the particular $(p,q)$ network that we consider below is
only one example of the sort of network these equations can describe.

\section{F- and D-String Network}

\begin{figure}[h] 
   \centering
   \includegraphics[width=3in]{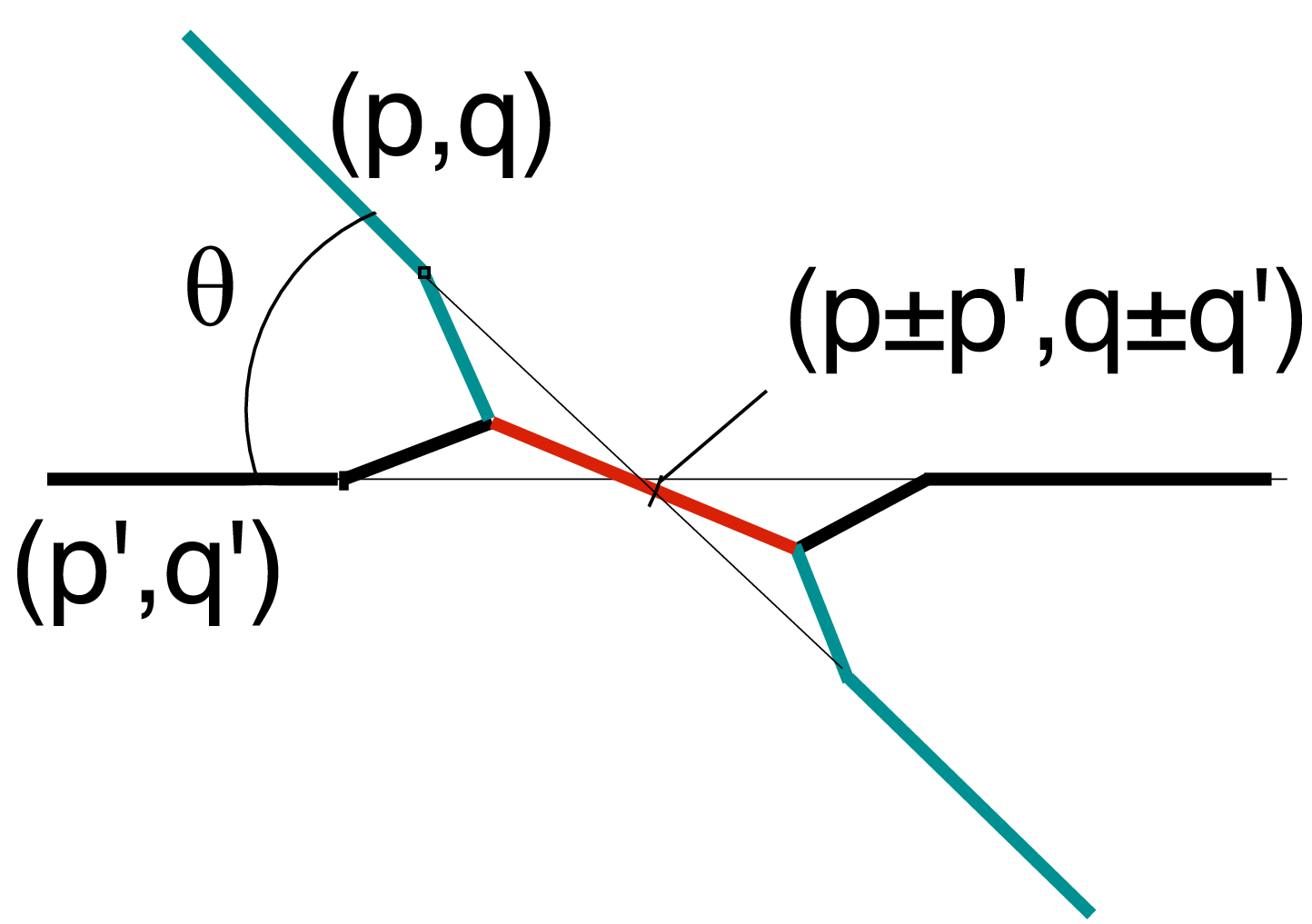} 
   \caption{A schematic view of a string intersection. The intersection angle, $\theta$, determines 
   whether the additive -- $(p+p',q+q')$ -- or subtractive -- $(p-p',q-q')$ -- binding occurs.
   \label{fig:interaction}}
\end{figure}

To specialize the preceding network to the $(p,q)$ strings of \cite{Copeland:2003bj,Jackson:2004zg},
 we define $P_{\alpha \beta \gamma}$ -- 
 taking $\alpha = (p,q)$, $\beta = (k,l)$, and $\gamma = (m,n)$  -- and
 motivate the overall interaction probability, $F$. For this investigation, 
 we make a first and very crude approximation: we assume that the 
 probability of two strings of different types interacting is a single, universal 
 constant, rather than a function of  of $\alpha, \beta, \gamma$ or the relative
  velocity of the strings. By discarding all these complexities, we retain
   only a kinematically determined branching ratio (see both Fig. \ref{fig:interaction} 
   and the discussion in the text below); 
   in future work, we may attempt to retain more of the physics contained in $F$ 
   to obtain more realistic results. Before we can write down this branching ratio, 
   we shall state the relevant properties of $(p,q)$ strings, since these determine the 
 form of $P_{\alpha \beta \gamma}$:
\begin{itemize}
\item Strings with positive and negative values of $p$ and $q$ are generically 
allowed; the sign of $p$ or $q$ indicates the direction of the string's charge. 
Because of a reflection symmetry, we can always choose the orientation
 of the string such that $p \geq 0$. For $p=0$, $q > 0$; for $p > 0$, $q \in \mathbb{Z}$. 

\item Strings with $q = 0$ are only stable for $p = 1$. It is probable that $(0,q)$ strings
are marginally bound; operationally, we assume that the non-zero momentum
transfers in the string collisions that accompany binding unbind these states. 
 An interaction that formally would create an $(N,0)$ or $(0,N)$ string thus, 
 in fact, creates $N$ $(1,0)$ or $(0,1)$ strings. 

\item We assume that two strings of different types interact with probability
 $F$. If two strings of different types do interact, there are two possible 
 products, or bound states, that they can form: a $(p,q)$ string interacting 
 with a $(p',q')$ string can form either a $(p+p',q+q')$ string or a $(p-p',q-q')$ string, 
 where we always take $p> p'$. As stated above, if either of these product 
 bound states has a resulting $p''$ and $q''$ that are not coprime, 
 then what is actually formed is a set of strings with stable, lower $p$, $q$ values.

 \item In agreement with our comments above, we assume $P_{0 \alpha \alpha} = 0$ or,
 equivalently, restrict our summations to $\beta \neq \gamma$.

\item For bound states of strings to be stable, $p$ and $q$ must be coprime.
If a bound state of a string is formed with $p$ and $q$ not coprime, then 
the new state is, in reality, a collection of lower-tension $p$ and $q$
strings that are coprime: i.e., a string which nominally has $p = Nk$, $q=Nl$ is actually
a set of $N$ $(k,l)$ strings. We may view this as the ``decay" of a $(Nk,Nl)$ string:
\begin{itemize}
	\item comes about because the $N$ $(k,l)$ strings that compose 
	this ``state" are BPS with respect to each other; 
	that is to say, they have no mutual binding interactions.  
	Any possible marginal binding may be ignored since the strings 
	are moving with relativistic speeds. 
	\item has no energy cost -- the resulting collection of strings always
	 has lower energy than the strings that bound to form them; there is
	 no actual $(Nk,Nl)$ bound state -- a collection of stable $(k,l)$ strings
	 is formed immediately following the interaction.
	\item can untie itself through reconnection events if the collection of
	 $(k,l)$ strings that are created in the interaction are tangled or tied
	  up immediately after their creation.
\end{itemize}
\end{itemize}

Whether the interaction of two strings forms a $(p+p',q+q')$ bound state
 or a $(p-p',q-q')$ bound state is determined by a simple consideration of 
 force balance: if the angle between the interacting strings is small enough,
 then the heavier, additive bound state is formed because the two interacting
strings' tensions can balance the tension of the heavier bound state; 
if the angle is greater than some critical angle of force balance, then 
the lighter, subtractive bound state is formed. The critical angle which determines 
which binding occurs is given by \cite{Jackson:2004zg}
$$
\cos \theta^{crit}_{klmn} = {e_{kl} \cdot e_{mn} \over |e_{kl}| |e_{mn}|} \;\;\;\; e_{mn} = ([m-Cn]g_s, n)
$$
where C is the RR scalar. If we assume a stochastic distribution of string orientations,
then the strings' interaction angle should have a flat distribution in $\cos \theta$, that is,
that each value of the cosine between -1 and 1 should be equally likely. If we assume this,
and remember that the directionality of the $F$ and $D$ charge must be taken into account --
i.e., $\theta = \pi/4$ is not equivalent to $\theta = 3 \pi / 4$ -- then the probability of forming the 
additive bound state is simply the fraction of cosine-space with $\theta$ less than the angle of 
force balance; the subtractive bound state is formed otherwise. We can write this as:
\be
P_{\alpha \beta \gamma}^{\pm} = \frac{1}{2} \left ( 1 \mp \left ( {(k+Cl)(m+Cn)g_s^2 + ln \over [(k+Cl)^2g_s^2 + l^2]^{1/2} [(m+Cn)^2g_s^2 + n^2]^{1/2}} \right ) \right ),
\label{pabc}
\ee
where $P^+$ indicates $\alpha = (p,q) = (k+m,l+n)$ and $P^-$
 indicates $\alpha = (p,q) = (k-m,l-n)$. This form 
captures the kinematic branching ratio, but as stated leaves out an important process: the 
creation of non-coprime $(p,q)$ strings, which are in reality collections of two or more 
coprime strings. To take this into account, we must slightly modify the 
way in which we insert  $P_{\alpha \beta \gamma}$ into our equations: we take
$$
P_{(Nk,Nl) (p,q) (p',q')} = N P_{(k,l) (p,q) (p',q')}.
$$
The inclusion of this process is extremely important. This break-up of non-coprime strings is
a nonreversible process that is fundamentally dissipative -- it helps to keep the average
tension of the network low both by limiting the pathways by which high-tension bound
states can be reached and by providing a mechanism through which a single interaction can 
destroy a high-tension bound state and replace it with a collection of low-tension strings.
Thus, in summary,
\begin{itemize}
\item Two strings of different types $\alpha$ and $\beta$ interact with probability $F$.
\item When these strings interact, either a subtractive or additive bound state is formed,
	with probabilities $P^{-}$ and $P^{+}$ given by Eq. (\ref{pabc}). 
	The two interacting strings are annihilated in the production of the new bound string state.
\item When the bound state $\gamma$ is stable, one such string is produced.
\item When the nominal bound state $\gamma$ is unstable, it immediately forms $N$ 
lower-tension stable strings, which helps keep the tension dependance of string density
spectrum steep.
\end{itemize}
In all our numerical runs we have assumed $C = 0$.

Some possible physical effects that our model neglects include:
\begin{itemize}
\item Velocity dependence of the interaction probabilities due to variation in the binding energy
of the resulting bound states: the increase binding energy that holds very high tension
states together is small -- the energy gained by binding decreases greatly for high-$\mu$ states. 
Thus we expect that the momentum transferred in even moderate-velocity interactions
that involve these lightly-bound states may lead them to unbind spontaneously.
\item This velocity dependence may lead to an effective cut-off in $\mu$, irrespective of 
the interaction dynamics.
\item We might not expect strings of widely varying tensions all to have the same velocities 
and characteristic length scale in reality, contrary to what our model assumes.
\item We have decoupled the evolution of the r.m.s. velocity and length scale of our network from 
the network's $P$ and $F$ dependent interactions, though one would generally expect
these interactions to be relevant to determining the network parameters.
\end{itemize}

\section{Network Results} 

\begin{figure}[ht] 
   \centering
   \includegraphics[width=3.5 in]{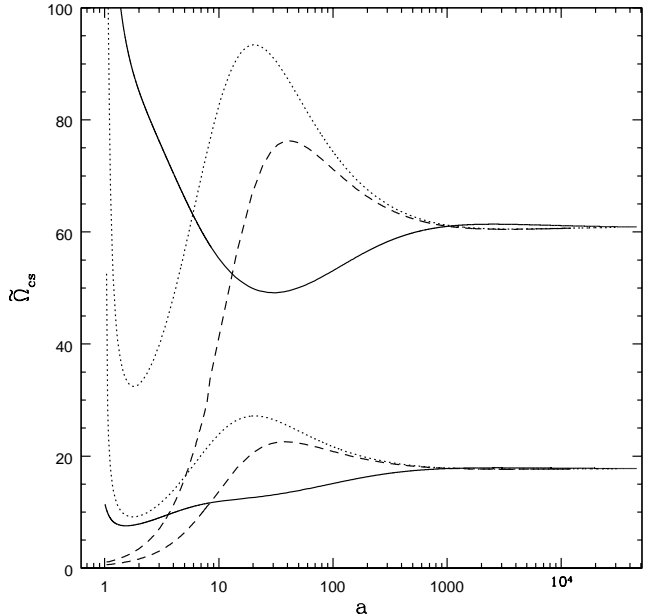}
   \caption{Comparison among three different sets of network initial conditions, all
   taking $F=1$, $P=0$. The higher three lines represent the evolution of the overall
   density in cosmic strings (summed over all string states),
    ${\tilde{\Omega}}_{cs} \equiv \Omega_{cs}/((8/3) \pi G\mu_{(0,1)})$,
   with scale factor, $a$. The lower three lines represent the evolution
    of the density in $(0,1)$, or $D$-, strings,
   ${\tilde{\Omega}}^{(0,1)}_{cs} \equiv \Omega^{(0,1)}_{cs}/((8/3) \pi G\mu_{(0,1)})$, with $a$.
    Our standard initial conditions, equal initial populations of $(1,0)$ and $(0,1)$ strings 
    and $HL = 1$, are shown by the dashed lines. The solid line
   represents the results from a network run with a short initial length scale ($10^{-2}$ of our 
   usual choice) and with over half of the initial string $(p,q)$ states in our network equally populated.
   Finally, the dotted line shows the results for a very large initial population of strings -- 
   ${\tilde{\Omega}}_{cs} \sim 1000$ --
   equally spread over half of the tension states included in our network, 
   with our usual choice for the initial network length scale.  For all the runs shown here we have
 set the superstring coupling, $g_s = 1.0$. 
   \label{fig:ICs} }
\end{figure}

The equations given in \S \ref{VOSmulti} require numerical solution. For all numerical results, 
we work in the radiation-dominated era, assume (for convenience) that the RR scalar, $C=0$,
and fix our constants $c_1=0.21$ and $c_2=0.18$ to match \cite{Martins} 
(these choices are made so that, at scaling, we have $HL = 0.1375$ and  $v = 0.655$). 
Furthermore we have done each run twice, with two different values of the superstring
coupling, $g_s=0.5$ and $1.0$.  We were
less certain about how to initialize the cosmic string network;  cosmic string creation
immediately after brane inflation is not understood completely.
Fortunately, scaling has proven to be quite robust
to a wide variety initial conditions; for an illustration of this, see Fig. \ref{fig:ICs}. On energetic
grounds, we believe that, in general, networks will be formed with primarily the lowest-lying states
populated; thus, for our calculations we chose initially to populate only the 
$(1,0)$ and $(0,1)$ states, and those with equal number densities. Final scaling results are always
insensitive to these choices; at worst, very different initial conditions
can alter the rate at which the network approaches the scaling regime.
We similarly find that any initial choices for network velocity, $v$, and length scale, $L$, quickly approach their analytically-predicted  scaling values. To integrate our equations 
numerically, another choice we had to make was how many
$(p,q)$ states to allow. After testing networks of many different sizes,
we found that our results showed a steep, power law
dependence of number density on tension in all cases. 
The relative densities of the low-lying tension states, furthermore,
were not changed when more high-tension states were included.
Thus we were able to obtain accurate results from a relatively small network: 
for the runs we show here, we have taken $p \in [0,5]$, $ q \in[-5, 5]$, though the 
way in which we solve the equations allows nominally higher-valued, temporary $(p,q)$
states to ``form" if the values of $p$ and $q$ are non-coprime, but only if the decay products
of the unstable $(p,q)$ state are a collection of stable $(k,l)$ strings with $k,|l| \leq 5$.
Finally, we take the scale factor of the universe $a=1$ at network initialization.

An interesting new result from this network model is that string networks with no loop creation --
those with $P = P_{0\alpha\alpha} = 0$ -- still exhibit cosmologically acceptable scaling; enough
energy is lost through string binding and binding-mediated annihilation to keep the comoving network
number densities $\Na \eta^2$ constant, regardless of initial conditions, after an initial relaxation period
following network formation. Because of this, these networks are very robust: though there are regions 
of parameter space where the network never truly reaches scaling, in all reasonable
cases (where we keep $F \neq 0$, $P \lesssim F$)  we never find 
cosmologically-disastrous solutions where cosmic strings come to dominate the energy density
of the Universe. 

There are three regimes of interest for these solutions:
\begin{enumerate}
\item For $F \gg P$ : The network will be dominated by $F$, $D$, and $(1,\pm1)$ strings. Higher tension states are present, but maximally suppressed (these networks have the steepest spectra). 

\item For $F \rightarrow0$: All string tensions that are present initially eventually reach the same scaling density. If there are a great many string states, this can cause a catastrophe, 
since the formation of loops tends to drive
all types of strings to the same value of $\Omega_{cs} / \mu_{\alpha}$.

\item For $P \sim F$: The interactions terms will populate the higher $(p,q)$'s, and the $P$ terms will flatten the spectrum somewhat because of its tendency to equally populate all levels. The larger $P$ is, the more quickly this happens. In Figs. \ref{omegafig}, \ref{multmu}, \ref{omegafigg05}, \& \ref{multmug05} a variety of combinations are shown. For larger values of $P$, the ``final" scaling state is not an exact scaling solution, but one that continues to evolve slowly to late times.
Some features that appear to generic in this regime include
\begin{itemize}
\item $\Omega_{cs}  / ( (8/3) \pi G\mu_{(0,1)}) = 60 /(F+0.55P)$ for $g_s = 1.0$, 
$46 / (F+0.53 P)$ for $g_s = 0.5$. This formula is only valid for $F\neq0$. 
\item $a_{\mbox{scaling}} \sim 1000$, so scaling is achieved at
 $T_{\mbox{scaling}} \sim 10^{-3} T_{\mbox{reheat}}$, where $T_{\mbox{reheat}}$ is the 
 temperature to which the universe reheats at the end of inflation. Since
 $T_{\mbox{reheat}} \sim M_{s}$, the string scale, $T_{\mbox{scaling}} \sim 10^{-3} M_s \gg \mbox{TeV}$,
 so scaling is reached long before the electroweak phase transition.
\item Scaling results are insensitive to initial conditions unless $F =0$
\item Steep final spectra: $N^{\mbox{final}}_{\alpha} \propto \mu_{\alpha}^{-n}$, $6<n \lesssim 10$
\end{itemize} 
\end{enumerate}

There are several aspects of these results that require further discussion. 
\subsection{Scaling}
\begin{figure}[h]
\centering
\includegraphics[width=3.5 in] {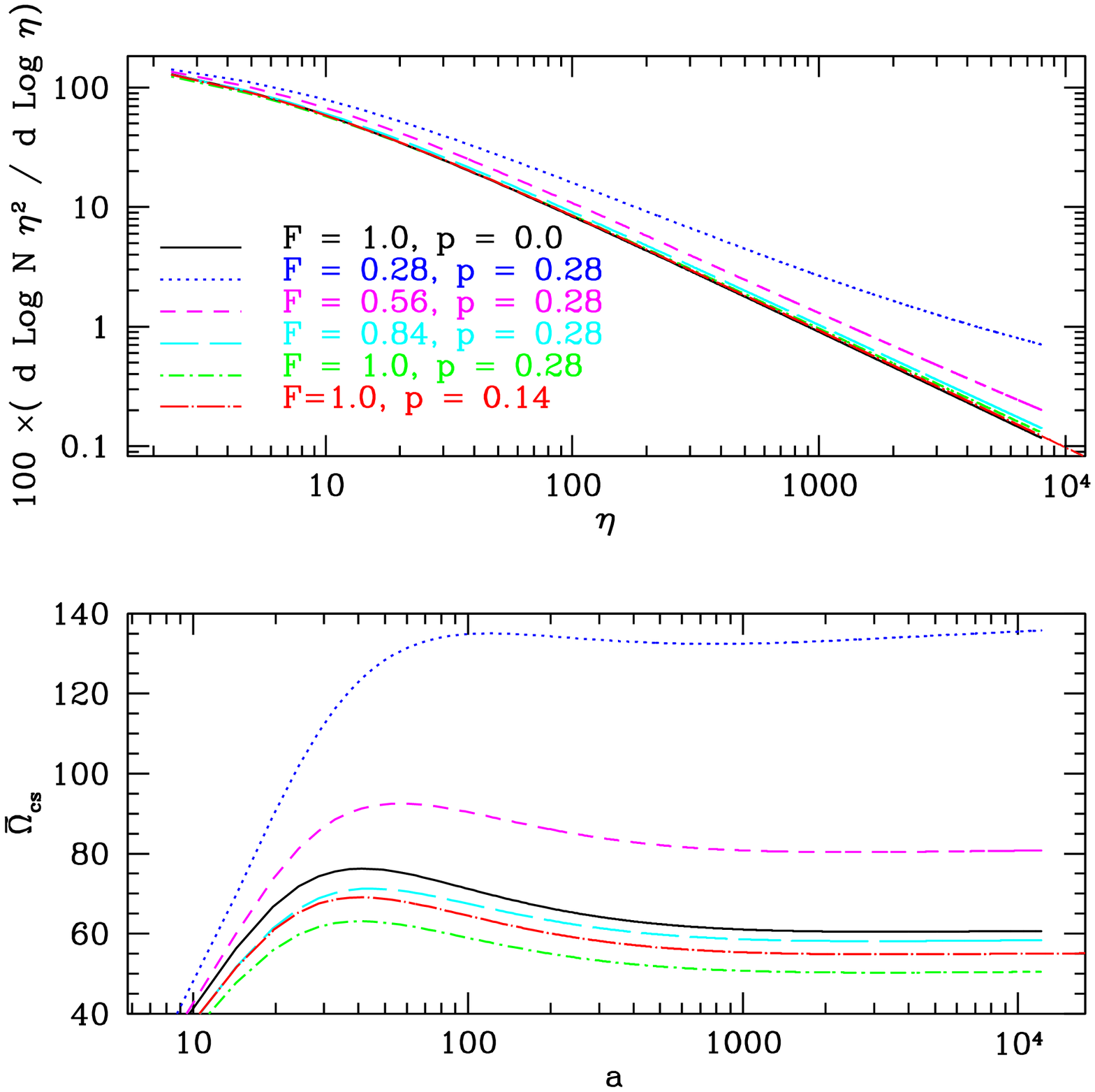} 
\caption{The bottom panel shows the evolution of 
$\tilde{\Omega}_{cs} = \Omega_{cs} / ((8/3) \pi G \mu_{(0,1)})$ 
for various parameter values, taking the string coupling $g_s=1.0$. The top panel shows the rate of change in the comoving number density $N \eta^2$; in the scaling regime, 
$d \log N \eta^2 / d \log \eta = 0$.    \label{omegafig}}
\end{figure}

\begin{figure}[h]
\centering
\includegraphics[width=3.5 in] {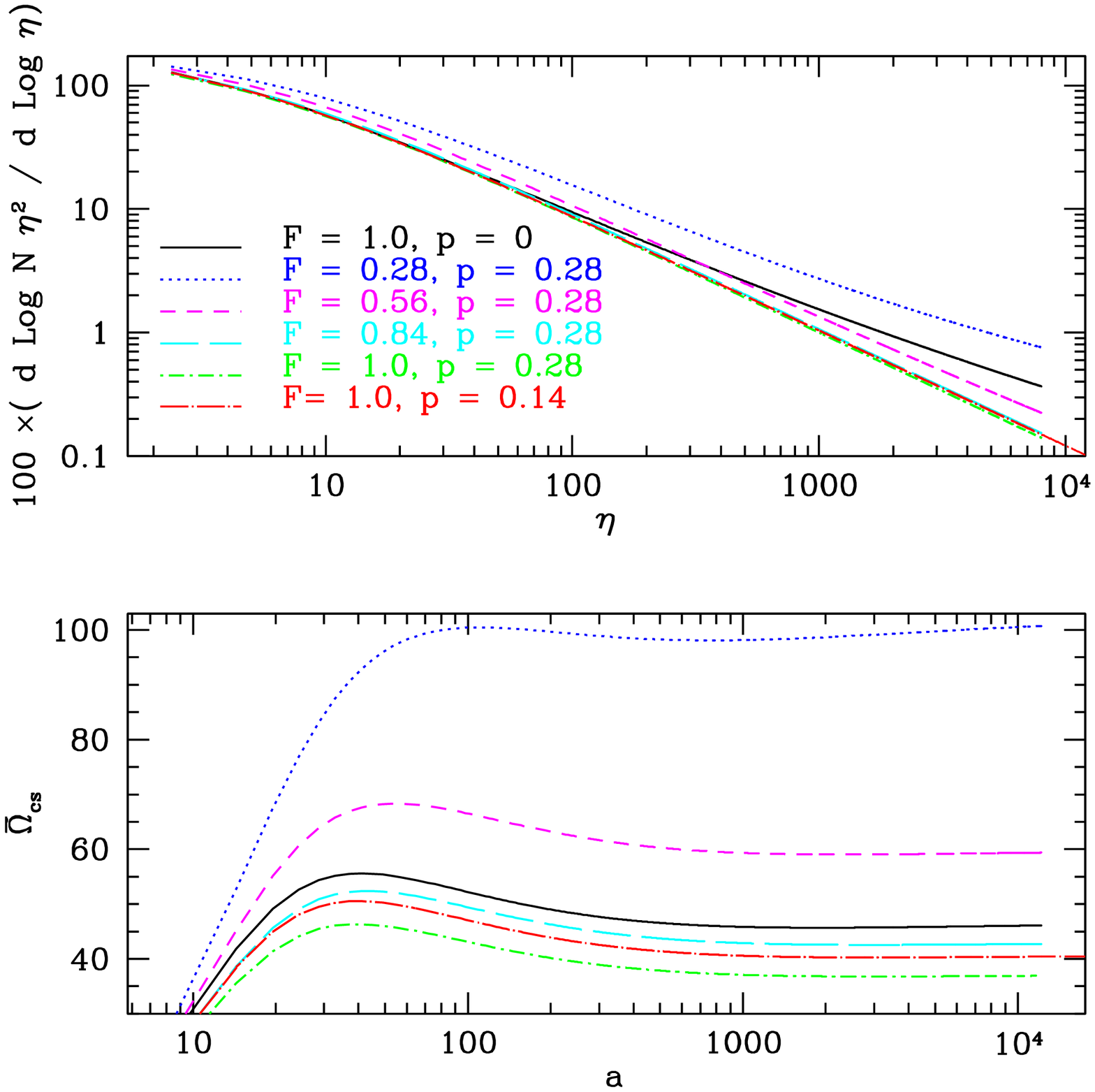} 
\caption{The bottom panel shows the evolution of $ \tilde{\Omega}_{cs} = \Omega_{cs} / ((8/3) \pi G \mu_{(0,1)})$
 for various parameter values, taking the string coupling $g_s=0.5$. The top panel shows the rate of
  change in the comoving number density $N \eta^2$; in the scaling regime,
   $d \log N \eta^2 / d \log \eta = 0$.    \label{omegafigg05}}
\end{figure}

\begin{itemize}
\item The overall properties of the networks are fairly insensitive both to initial conditions 
\cite{barnaby} and to particular parameter choices; i.e., $\Omega_{cs}$ never grows fast 
enough ever to come to dominate the universe.
\item However, the final state of the string network depends upon the relationship between $F$ and
$P$. When $P \sim F$, the network does not quickly reach a true scaling solution. Instead, it 
continues to evolve to late times (see Figs. \ref{omegafig} and \ref{omegafigg05}).
Because of the efficient energy loss from both binding 
and loop formation, the network's overall density grows very slowly --
e.g., $(d \log \Omega / d \log a) \sim 0.07$ for $a \sim 10^4$; $\sim 0.01$ for $a \sim 10^5$, for 
the case of $P = F = 0.28$;  
note that $(d \log \Omega / d \log a) = 0$ defines entry into the scaling regime.
This late growth comes about because of the continuing competition between loop formation,
which wants equally to populate all string states, and binding interactions, which tend
to destroy high-tension bound states. This late evolution is not dangerous cosmologically.
\item In agreement with our understanding of the late evolution of $P \sim F$ networks, such
networks tend to develop somewhat flatter final spectra, as their high-tension bound states
tend to be more populated than those in $F \gg P$ networks. Their spectra still exhibit very 
steep power law behavior, however (see Figs. \ref{multmu} and \ref{multmug05}).
\item The scale factor at which the network enters the scaling regime is somewhat dependent 
upon initial conditions: networks with more states initially populated tend to take slightly longer to reach 
scaling, though the greater frequency of interactions caused by such initial conditions means
that these networks tend to be less dense throughout their evolution than 
networks that are formed with only low-lying string states (see Fig. \ref{fig:ICs}); networks
that begin with much smaller initial $L$, on the other hand,
can take a good bit longer to reach scaling. 
Recall also that the binding interactions of high-tension states will very often lead to
non-coprime combinations, which leads such networks quickly to develop the same 
kinds of steep spectra that are seen when less democratic initial conditions are used.
\end{itemize}
\subsection{Low-F Catastrophe}
\begin{itemize}
\item An aspect of superstring networks that has been as-yet unappreciated is that their
ability to populate arbitrarily high tensions through the formation of bound
states can lead to a cosmological catastrophe if such states cannot be made to decay. 
In traditional network evolution, which is what our equations reduce to 
when $F \rightarrow 0$, even very small initial populations of each possible string state will
each eventually reach the same final scaling density. When this happens, 
$\Omega_{cs} = \sum_{\alpha} \Omega^{cs}_{\alpha} \propto \sum_{\alpha} \Nalpha \mu_{\alpha}$
can become huge, and thus disastrous,
even if the energy density in each individual state is small.
\end{itemize}

\subsection{Final Spectra}
\begin{figure}[h] 
   \centering
   \includegraphics[width=3.5 in]{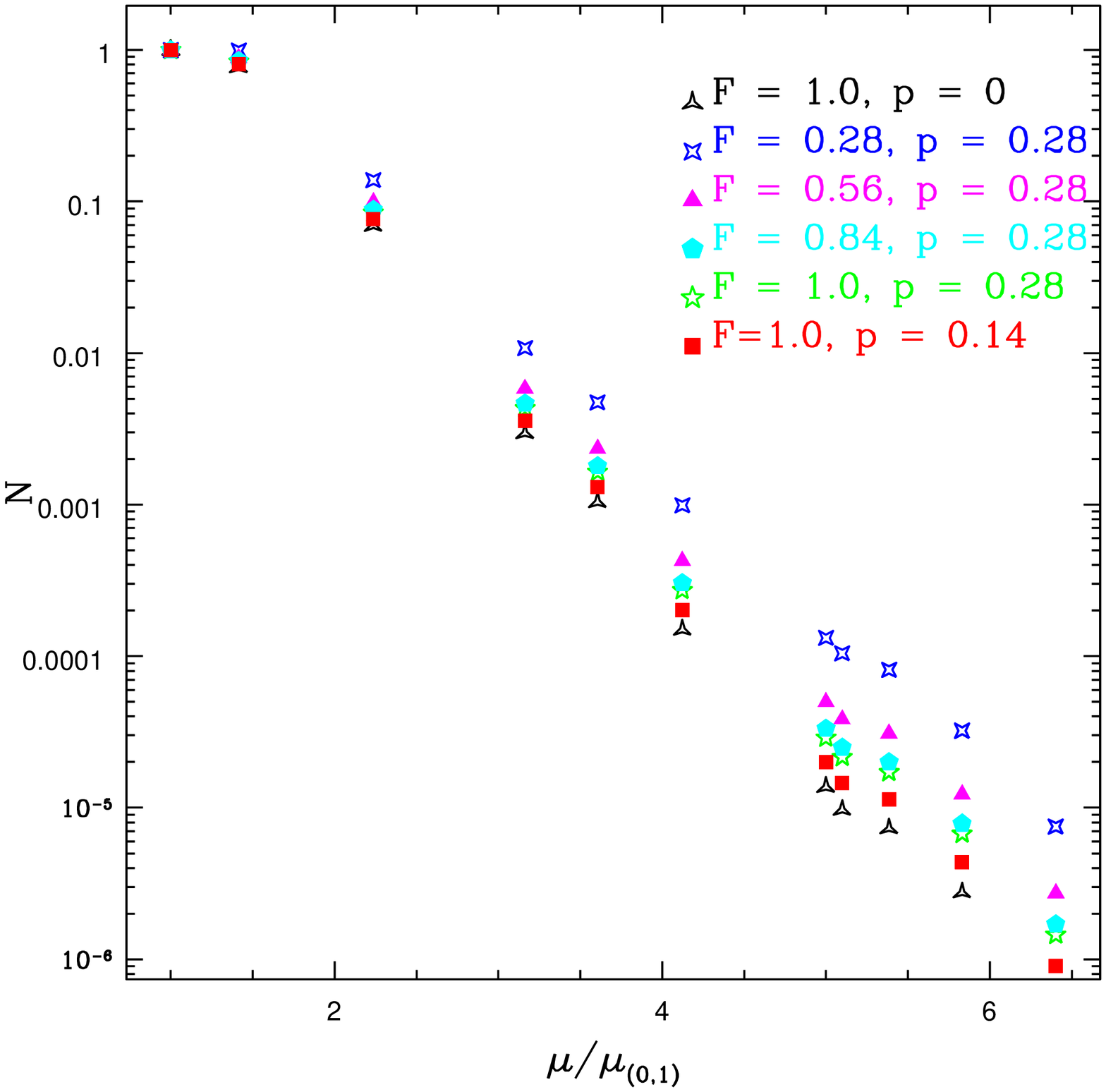} 
   \caption{The final scaling-era spectra for a variety of parameter combinations, taking the string
   coupling $g_s=1.0$, with $N_{(0,1)}$ normalized to unity and the other number densities
  altered accordingly.    \label{multmu}}
\end{figure}

\begin{figure}[h] 
   \centering
   \includegraphics[width=3.5in]{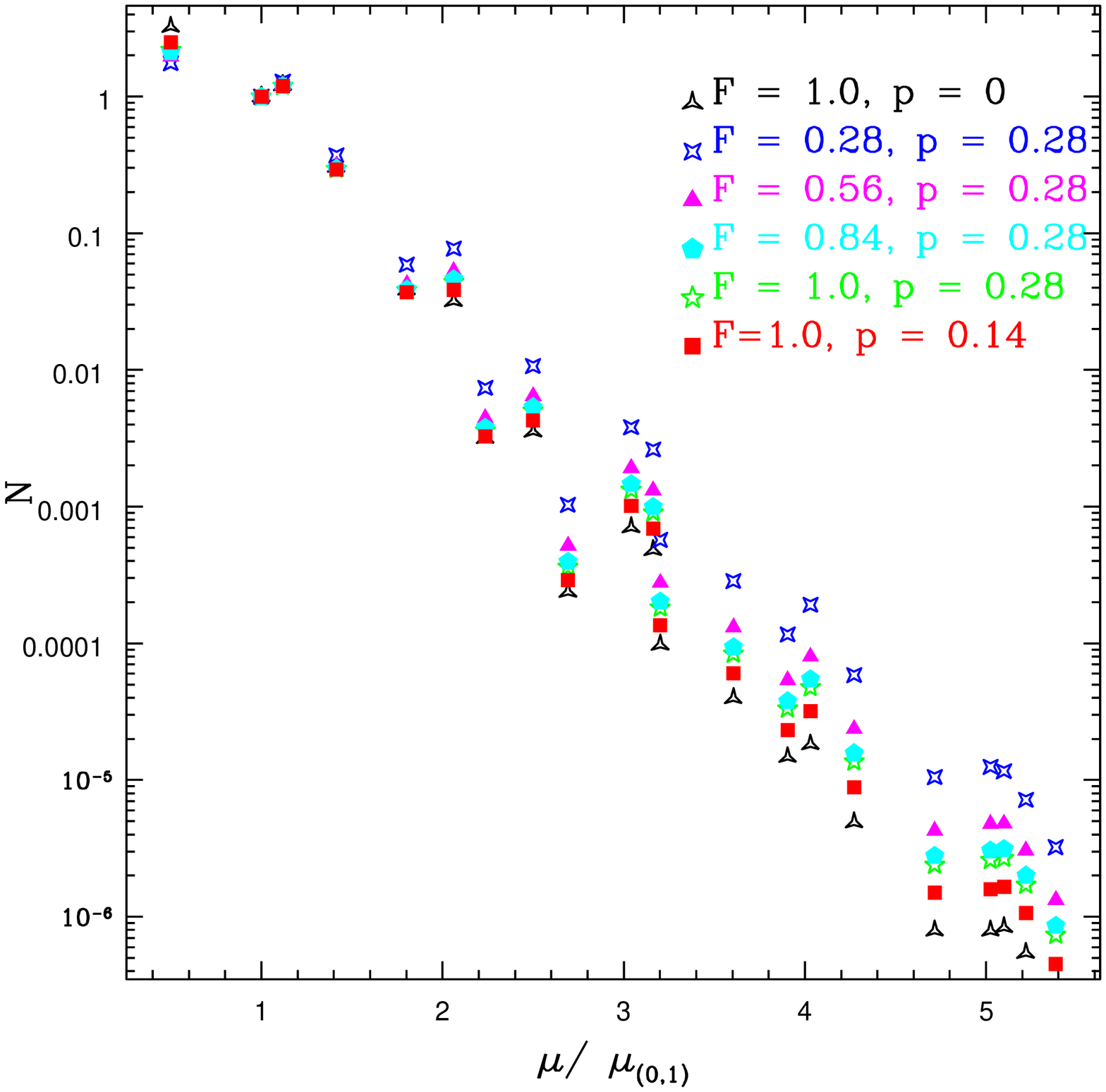} 
   \caption{The final scaling-era spectra for a variety of parameter combinations, taking the string
   coupling $g_s=0.5$, with $N_{(0,1)}$ normalized to unity and the other number densities
   altered accordingly.    \label{multmug05}}
\end{figure}

\begin{itemize}

\item The fact that our numerical solutions have found a very strong dependence of string
number density on tension -- with $N_{\alpha} \propto \mu_{\alpha}^{-n}$, and $6 < n \lesssim 10$
-- is another important aspect of these networks. If the spectrum were flat, or nearly so, 
a scenario very much like the low-F catastrophe outlined above would ensue: since
the effect of many string states is additive, and since there are many more possible states
at higher tensions, such a flat-spectrum network would be ruled out immediately
by cosmological considerations. 
\item Computationally, the steepness of the spectra greatly eases our task. Numerical 
tests showed that the addition of many high-tension states with low number densities
scarcely affected any of the results. We were thus able to limit ourselves to small networks,
with $p_{\mbox{max}} = |q_{\mbox{max}}| = 5$. 
\item Careful study of Figs. \ref{multmu} and \ref{multmug05} shows 
that the spectra plotted there are not strict power laws. The spectra are, in fact, somewhat
flatter for very low tensions, where the number of possible states is small. Thus, in all 
cases there are proportionally more $F$, $D$, and $(1,1)$ strings than anything else; 
particularly when $F \gg P$, these states will dominate the cosmic string network. The relative
populations of these low-lying states are tabulated in Tab. \ref{tab:ratios}. 
\item The effect of varying the superstring coupling, $g_s$, is to vary the tension of the F
strings relative to the D strings, with this variation propagating up the ladder of bound
states. Here, we have taken only two representative values of $g_s$: 0.5 and 1.0. 
Reducing $g_s$ affected the network as our $N \propto \mu$ results suggest:
those states which were previously degenerate (equally populated)-- 
$(1,0)$ or $[(2,1)+(2,-1)]$ relative to $(0,1)$ or  $[(1,2)+(1,-2)]$, for instance --
had their degeneracy lifted, with the lighter state's number density increasing. The precise amount 
of increase was dependent upon the values of $F$ and $P$. Again, see Tab. \ref{tab:ratios}.
\item We note in passing that (unphysical) networks with $P = 0$ and with the loop term
$\propto c_2$ removed from Eq. (\ref{popeqn}) also go to scaling.
\item In all our networks, the lowest tension states dominate the network energy density. We
expect this to be a feature of any multi-tension networks that interact via binding, even if
the spectrum of possible bound states is much more complicated than the one we have considered.
\end{itemize}

\begin{table}
\begin{center}
\begin{tabular}{|c|c|c|c|c|c|}
\cline{3-6}
\multicolumn{2}{c}{} \vline& \multicolumn{2}{c}{$g_s = 1.0$} \vline &  \multicolumn{2}{c} {$g_s = 0.5$} \vline \\
\cline{3-6}
\multicolumn{2}{c}{}  \vline & \multicolumn{2}{c}{${\mu_{(1,0)} \over{ \mu_{ (0,1)}}} =  1\;$  ${\mu_{(1,\pm1)} \over { \mu_{ (0,1)}}}= \sqrt{2}$}  \vline &   \multicolumn{2}{c}{${\mu_{(1,0)} \over {\mu_{ (0,1)}}} = {1\over2} \;$   ${\mu_{(1,\pm1)} \over{ \mu_{ (0,1)}}} = 1.12$}  \vline \\
\hline
F & P & $N_{(1,0)}/N_{(0,1)}$ &$N_{(1,\pm1)}/N_{(0,1)}$  & $N_{(1,0)}/N_{(0,1)}$ & $N_{(1,\pm1)}/N_{(0,1)}$ \\
\hline
1 & 0 & 1 & 0.769 & 3.24 & 1.21 \\
1 & 0.14 & 1 & 0.803 & 2.50 & 1.19\\
1 & 0.28 & 1 & 0.836 & 2.19 & 1.19 \\
0.56 & 0.14 & 1& 0.829 & 2.24 & 1.19  \\
0.56 & 0.28 & 1 & 0.887 & 1.96 & 1.21\\
\hline
\end{tabular}
\end{center}
\caption{\label{tab:ratios}The relative populations of the three
 lowest-lying tension states, which in all cases dominate the networks' energy density.}
\end{table}


\section{Observational Consequences}

Several aspects of $(p,q)$ networks are potentially observationally distinct from regular cosmic string networks. The most obvious difference is that these networks feature a spectrum of string tensions. 
We suggest a few possible observational signatures that could allow one to distinguish
a $(p,q)$ network from a standard, abelian Higgs network:

\begin{itemize}

\item Previous studies of cosmic string lensing probability \cite{hogan} have been 
based on results from standard, abelian Higgs network models. In such networks,
$\Gamma = \Omega_{cs} / (8\pi G\mu) \approx 20$. In principle, our model allows values
for $\Gamma$ both less than and greater than the abelian Higgs value. However, 
we expect the extra-dimensional nature of superstrings to reduce their interaction
rates, which leads to higher values
of $\Gamma$ and $\Omega_{cs}$.
If future lensing surveys find a rate of cosmic string 
lensing substantially higher than that predicted by a abelian Higgs network, 
that rate could both be a signature of 
a cosmic superstring network as well as an observational constraint
on the parameters of such a network. 

\item If the overall densities in cosmic superstring networks are generally higher
than those in abelian Higgs models, then observational bounds on cosmic
string tension (e.g. \cite{us05}) that depend on overall network properties 
will need to be reinterpreted. We expect that the net effect will be to tighten
such bounds, though how much the bounds will change is difficult to predict 
since observational bounds depend on many aspects of string networks (e.g., 
string substructure, or ``wiggliness"), while the detailed properties of 
multi-tension networks have not yet been fully fleshed out.

%
 
\item The $Y$-shaped junction of two strings in the act of binding is a good signature of the 
non-trivial properties of cosmic superstrings. Such a junction, if present, 
could be detected by cosmic string lensing, or by observation of the Kaiser-Stebbins effect,
 where a temperature difference is seen in the cosmic microwave background radiation 
due to a string-induced Doppler shift  \cite{deltat}. 
In the latter case, we would expect to see a different temperature in each of the 3 patches of sky. 
The relativistic motion of the binding strings could also be an indicator of a binding event:
the cosmic string lensing angle is enhanced by 
a factor of $\gamma$ for moving strings \cite{Shlaer:2005gk}
(depending on string orientation), which is 
moderate for usual network motions ($\sim1.3$, in the radiation-dominated era).
 The strings motions near a binding site, however,
are very relativistic, though over a very small spatial region, and thus would exhibit
exaggerated lensing near the binding site.  Random variation among string
velocities within the network should also lead to the existence of some 
individual fast-moving strings whose lensing will also be enhanced.

\item A recent analysis \cite{wright} of the direct detectability of cosmic string-generated
temperature anisotropies in the data from the upcoming Planck satellite suggests
that relatively high tension or fast moving strings within cosmic superstring networks
would be marginally within the Planck range of detectability (they estimate that 
strings with $G \mu \approx 6 \times 10^{-6}$, $v = 1/\sqrt{2}$ would be directly detectable by 
Planck; for a multi-tension network with a fiducial tension $G \mu \sim 5 \times 10^{-7}$,
only strings with a combined $\beta \gamma$ and high-$(p,q)$ tension enhancement 
of $\sim 10$ would be seen).

\item
Direct observation of more than one cosmic string tension from observational techniques, such as 
gravity wave bursts \cite{damour} or gravitational lensing, that are sensitive to a particular 
string's tension would be a definite prediction of this kind of network. In the case of
lensing, however, the velocity, orientation, and string substructure dependences 
of the lensing angle may overwhelm this effect for the most probable 
lensing strings (since over 90\% of the strings in our network 
are $F$, $D$, and $(1,\pm1)$ strings, whose tensions are all of the
same approximate magnitude). For random string orientations, the
string lensing angle can vary by as much as a factor of six because of
velocity dependent effects, though we expect typical velocity 
dependent variation of only a factor of two or so (these variations arise 
because the string lensing angle is proportional both to the sine of the orientation angle
of the cosmic string relative to the line of sight as well as $\gamma (1 + \hat{n}\cdot v)$,
where $v$ is the string velocity and $\hat{n}$ is a unit vector in the direction between
the observer and the string \cite{Shlaer:2005gk}).
Thus, there are two ways that lensing measurements could indicate the existence 
of a multi-tension network. The most dramatic would be  a single very large
 ($\gtrsim10$) variation between two observed lensing angles. Just as compelling, however,
 would be if a large number of lensing measurements were made with a typical
variation among events that is greater than one would expect  to arise, statistically,
from random string orientations and string velocity directions. 
In any event, accurate follow-up observations of the Kaiser-Stebbins 
effect, where the string's relative velocity enters the equations differently (in a cross
product rather than a dot product) could perhaps allow us to disentangle to some degree
the string's velocity and intrinsic tension. Another possible avenue for discriminating between
variation due to string orientation and velocity and intrinsic string tension would be if a 
series of lensing events were observed along a single long string within a small patch of the sky.
Since we expect strings to be curved, it could be possible to observe the same cosmic string
at several different orientations. This could allow us to extract the string's actual tension
through a statistical analysis of the events' lensing angles.

\item
Another signature would be a mismatch between a particular string tension
measured directly -- from lensing, perhaps -- and a $G\mu$ measurement
coming from a technique like CMB fluctuations or pulsar timing analysis
that is only sensitive to the averaged network as a whole; 
however, the effects of string substructure 
(i.e., string wiggliness), which alter string-generated CMB spectra,
could mask more subtle expressions of this effect.  If limits on 
string substructure (see \cite{us05}) improve, then a direct detection via lensing
of a string with a tension that is several times larger than what CMB limits would lead us 
to expect for a single-tension network would be a strong indication of the 
existence of a multi-tension network.
  

\end{itemize}

By combining different observations and accumulating sufficient data, one should be able to measure a set of properties of the cosmic strings and so distinguish between different scenarios. This goal will be easier to reach if the true cosmic string tension is closer to today's observational bound.
%

\vspace{6mm}

{\bf Acknowledgments}

\vspace{3mm}

We thank Nick Jones and
David Chernoff for valuable discussions. This material is based upon work supported by the National Science Foundation under Grant No. PHY-009831. M.W. is supported by the NSF Graduate Fellowship.

\end{document}